\documentclass{webofc}

\usepackage{hyperref}

\usepackage[varg]{txfonts}   
\usepackage{tikz}
\usepackage{bm}
\usepackage{graphicx}   
\usepackage{amsmath,amssymb,mathrsfs}  
\usepackage[utf8]{inputenc}
\usepackage{color}

\def\tr{\mathrm{tr}}

\def\mn{{\mu\nu}}
\def\llangle{\left\langle}
\def\rrangle{\right\rangle}

\begin{document}
\title{Transport near the chiral critical point
}

\author{
        \firstname{Alexander} \lastname{Soloviev}\inst{1}\fnsep\thanks{\email{alexander.soloviev@stonybrook.edu}} 
}

\institute{Center for Nuclear Theory, Department of Physics and Astronomy,
Stony Brook University, Stony Brook, New York 11794, USA
          }

\abstract{%
 The evolution of a heavy ion collision passes close to the O(4) critical point of QCD, where fluctuations of the order parameter are expected to be enhanced. Using the appropriate stochastic hydrodynamic equations in mean field near the the pseudo-critical point, we compute how these enhanced fluctuations modify the transport coefficients of QCD. Finally, we estimate the expected critical enhancement of soft pion yields, which provides a plausible explanation for the excess seen in experiment relative to ordinary hydrodynamic computations.
}
\maketitle



\section{Introduction} \label{intro}
Relativistic viscous hydrodynamics \cite{Jeon:2015dfa} is remarkably successful at describing measurements performed at the Relativistic Heavy Ion Collider (RHIC) and the Large Hadron Collider (LHC). One feature present in heavy ion collisions, yet absent from current hydrodynamic simulations, is chiral symmetry breaking and the associated chiral phase transition. Although chiral symmetry is not an exact symmetry due to the finite quark mass, for small enough quark mass the pattern of chiral symmetry breaking should have some influence on the evolution of the system. This is supported by results from the lattice \cite{HotQCD:2019xnw,Kaczmarek:2020sif}, where certain aspects of QCD thermodynamics, such as chiral susceptibility, could be understood using $O(4)$ scaling functions. As such, chiral symmetry and its breaking should be included to increase the predictive power of hydrodynamics.

The reorganization of hydrodynamics to include chiral physics was presented recently in \cite{Grossi:2020ezz,Grossi:2021gqi}, building on earlier work on $O(4)$ physics going back to Rajagopal and Wilczek \cite{Rajagopal:1992qz} almost thirty years ago. The basic idea is that at long distances and high temperature, the appropriate effective theory for QCD is ordinary hydrodynamics, approximately respecting the $SU_L(2) \times SU_R(2)\sim O(4)$ symmetry. As the temperature drops below the critical temperature, the system undergoes the chiral phase transition and develops a chiral condensate, breaking the symmetry down to $SU_V (2)$. Thus, the ordinary hydrodynamic theory, described by energy-momentum and charge conservation, should be expanded to include the fluctuations of the order parameter, e.g. the pions, which are the pseudo-Goldstone modes associated with the broken chiral symmetry. The resulting theory is reminiscent of a non-abelian superfluid \cite{Son:1999pa,Son:2002ci}.




 In these proceedings, we review the work of \cite{Grossi:2021gqi}, where we computed the contribution of the pseudo-Goldstone modes to the hydrodynamic transport coefficents, namely the shear viscosity and iso-vector conductivity. 
We will work in the mean field limit to attain a qualitative understanding of the scaling functions, while noting that recently the real time dynamics of the critical theory near the $O(4)$ critical point have been studied \cite{Florio:2021jlx}. 

 
The organization of the paper is as follows: first, we provide a recap of the theoretical set up in Sec.~\ref{sec:setup}. Then in Sec.~\ref{sec:mod} we outline the  computation of the modification of the transport coefficients (here we discuss the shear viscosity. For the isovector conductivity, see \cite{Grossi:2021gqi}). Finally, we conclude in Sec.~\ref{sec:outlook} by making a few phenomenological remarks on the expected soft pion enhancement compared to hydrodynamic models which do not include chiral effects.


\section{Hydrodynamics near the O(4) critical point}\label{sec:setup}

We begin with the generating functional
\begin{align}
W[g_\mn, A_\mu, \phi_a] &=\int d^3 x\; \left[ {p(T)}+\frac{\chi_0}{4}\mu_{ab}^2-\frac{1}{2}\Delta^{\mu\nu}D_\mu\phi_a D_\nu\phi_a-V(\phi)\right]
,
\end{align}
where $p$ is the pressure, $\mu_{ab}$ is the $O(4)$ chemical potential (e.g. $\mu_{0i}=\mu_A$), the $O(4)$ vector is $\phi_a=(\sigma, \pi_i)$ with the $\sigma$ denoting the condensate and the $\pi^i$ are the pions, the gauge covariant derivative is
$D_\mu \phi= \partial_\mu \phi-iA_\mu \phi/2$, the potential is $V(\phi_\alpha \phi_\alpha)=m^2_0 (t) \phi_a \phi_a+\lambda (\phi_a\phi_a)^2/4-H_a\phi_a$,
with the mass $m^2_0(t)\sim (T-T_c)\sim z$ and $H_a$ is the applied magnetic field or quark mass, i.e. the explicit symmetry breaking term. 
We  define the temperature via
\begin{align}
T=\frac{1}{\sqrt{-\beta^\mu g_\mn \beta^\nu}}\, ,
\end{align}
where $\beta^\mu=u^\mu/T$ and $u^\mu$ is the normalized four velocity, $u^\mu u_\mu=-1.$ Note that $a$ and $b$ denote the $O(4)$ index, Greek indices denote spacetime indices and $i$, $j$ and $k$ are spatial indices.


We compute the energy momentum tensor and the current via
\begin{align}
T^{\mu\nu}&=\frac{2}{\sqrt{-g}}\frac{\delta W}{\delta g_{\mu\nu}}, \quad
J^\rho_{ab}=\frac{2}{\sqrt{-g}}\frac{\delta W}{\delta (A_{\rho})_{ab}}.
\end{align}
The former is exactly conserved, while the latter is only partially conserved
\begin{align}
 \partial_\mu T^{\mu\nu}=0, \quad \text{and} \quad 
\partial_\mu  J^\mu_{ab}={\phi_a H_b-\phi_b H_a}.
\end{align}
At ideal order, we should check that entropy, given by the Gibbs-Duhem relation,
\begin{align}
s=\frac{\varepsilon+p-\frac{1}{2}\mu\cdot n}{T},
\end{align}
is not being produced. We find that the entropy is conserved up to a term on the right hand side
\begin{align*}
\partial_\mu (su^\mu)=\frac{1}{T}
{\left[u\cdot\partial \phi_a+{\mu_{ab}\phi_b}\right]}
\left[\partial_{\mu}(\Delta^{\mu\nu}\partial_\mu\phi_a)-\frac{1}{\phi^2}\frac{\partial V}{\partial \phi^2}\phi_a+H_a)\right] \, .
\end{align*}
Requiring ideal entropy production to be zero, we see immediately the Josephson constraint
\begin{align}
u\cdot\partial \phi_a+{\mu_{ab}\phi_b}=0,
\end{align}
which encodes the coupling between the pions and the chemical potential.

Next we add dissipation via
\begin{align}
T^{\mu\nu}&=T^{\mu\nu}_{\rm ideal}+\Pi^{\mu\nu}\\
J^\mu&=J^\mu_{\rm ideal}+q^\mu \\
u\cdot\partial \phi_a&+{\mu_{ab}\phi_b}=\Xi_a
\end{align}
The requirement of positive entropy production leads to dissipative equations \cite{Grossi:2021gqi}, namely the relaxation equation for the order parameter
\begin{align}
{u^\mu \partial_\mu \phi_a+\mu_{ab}\phi_b}={\Gamma \left[\partial_{\mu}(\Delta^{\mu\nu}\partial_\mu\phi_a)
-\frac{1}{\phi^2}\frac{\partial V}{\partial \phi^2}\phi_a+H_a)\right] 
+ \zeta^{(1)}\phi_a \nabla_\alpha u^\alpha}
\end{align}
and the partially conserved axial current ($\partial_\mu J^\mu_{ab}=\phi_a H_b-\phi_b H_a$)
\begin{align}J^\mu_{ab}=
{n_{ab}u^\mu+\Delta^\mn(\phi_b (D_\nu \phi)_a-\phi_a (D_\nu \phi)_b)}
{- T\sigma\Delta^\mn\partial_\nu \left(\frac{\mu_{ab}}{T}\right)} \, ,
\end{align}
For completeness, the conserved energy momentum tensor ($\partial_\mu T^\mn=0$) reads
\begin{align}
T^{\mu\nu}&={(\varepsilon+p)u^\mu u^\nu+p g^{\mu\nu}+(\partial_\mu \phi_a)^2
-u^\mu u^\nu (u^\sigma \partial_\sigma \phi_a)^2}\nonumber\\
&{-\eta \sigma^{\mu\nu}\label{emt}
-\Delta^\mn\zeta^{(0)}\nabla\cdot u} {+\Delta^\mn\zeta^{(1)}\phi_a
\left[\partial_{\mu}(\Delta^{\mu\nu}\partial_\mu\phi_a)
-\frac{1}{\phi^2}\frac{\partial V}{\partial \phi^2}\phi_a+H_a)\right]\, ,
}
\end{align}
where the first line and second line are the ideal and dissipative contributions, respectively.

As mentioned in the introduction, we will work in the mean field limit with $\phi_a=(\sigma,\bf{0})$ and $H_a=(H,\bf{0})$, which is found by minimizing the potential
\begin{align}\label{min-pot}
0=\frac{dV}{d\phi}=m_0^2(t) \sigma+\lambda \sigma^3 -H.
\end{align}
We will call the solution of the above equation $\bar\sigma=\bar\sigma(T,H).$ We can compare our scaling function in the mean field to the one computed on the lattice \cite{Engels:2009tv, Engels:2011km} and find that there is close qualitative agreement.
Next, we linearize our equations around mean field, $\phi_a\rightarrow \phi_a+\delta \phi_a=(\bar{\sigma}+\delta \sigma, \bar{\sigma}\varphi_i),$ which leads to the following set of equations:
\begin{align}\label{lin-eom}
\partial_t\varphi&=
-{\mu}_A
+{\Gamma(\nabla^2 -m^2)\varphi},\\
{\partial_t \mu_A}&= {v^2(-\nabla^2+m^2)\varphi}
+{D_0 \nabla^2 \mu_A},\\
\partial_t \delta\sigma&={\Gamma (\nabla^2-m_\sigma^2)\delta\sigma}.
\end{align}
Parameters here depend on the mean field \eqref{min-pot} and temperature
\begin{align*}
v^2=\bar{\sigma}/\chi, \quad \quad m^2&=H/\bar{\sigma}, \quad \Gamma=\text{const}, \quad D_0=\text{const}, \\
   m_{\sigma}^2 &= m_c^2 \left(z + 3 f_G^2(z) \right).
\end{align*}

We can compute the propagators by going to Fourier space, adding a fictitious source and solving the resulting equation. As an example, the diffusive equation for the condensate reads
\begin{align}
-i\omega +\Gamma (k^2+m_\sigma^2)\delta\sigma=\Gamma H, 
\end{align}
where $H$ is the source. Then the retarded propagator is simply
\begin{align}
G_R^{\sigma\sigma}=\frac{\Gamma}{-i\omega +\Gamma (k^2+m_\sigma^2)}.
\end{align}
To find the symmetrized propagator, we simply need
\begin{align}
G_{\rm sym}=\frac{T}{i\omega}( G_R-G_A),
\end{align}
where the advanced propagator is simply the complex conjugate of the retarded propagator, $G_A=(G_R)^*.$ The complete set of symmetric propagators is found to be
\begin{align}
G_{\rm sym}&=\nonumber
\begin{pmatrix}
G_{\rm sym}^{\varphi\varphi} & G_{\rm sym}^{\varphi\mu} & G_{\rm sym}^{\varphi\sigma}\\
G_{\rm sym}^{\mu\varphi} & G_{\rm sym}^{\mu \mu} & G_{\rm sym}^{\mu \sigma}\\
G_{\rm sym}^{\sigma\varphi} & G_{\rm sym}^{\sigma\mu} & G_{\rm sym}^{\sigma\sigma}\\
\end{pmatrix}\\
&=
\begin{pmatrix}
\frac{2T}{\chi_0}\frac{g_1(\omega^2+g^2_2)+g_2^2\omega_k^2}{(-\omega^2+\omega_k^2+g_1 g_2)^2+(\omega \Gamma_k)^2} 
&\frac{2T}{\chi_0}\frac{-i\omega_k \omega \Gamma_k}{(-\omega^2+\omega_k^2+g_1 g_2)^2+(\omega \Gamma_k)^2} & 0\\
\frac{2T}{\chi_0}\frac{i\omega_k \omega \Gamma_k}{(-\omega^2+\omega_k^2+g_1 g_2)^2+(\omega \Gamma_k)^2} & \frac{2T}{\chi_0}\frac{g_2(\omega^2+g^2_1)+g_1^2\omega_k^2 }{(-\omega^2+\omega_k^2+g_1 g_2)^2+(\omega \Gamma_k)^2}&0\\
0 & 0& \frac{2T\Gamma}{\omega^2+\Gamma^2(k^2+m_\sigma^2)}\\
\end{pmatrix}\, ,
\label{propagator}
\end{align}
where we used the following shorthand
\begin{align}
g_1&=\Gamma(k^2+m^2), \quad 
g_2 =D k^2, \quad
\Gamma_k=g_1+g_2, \quad
\omega_k=v^2(k^2+m^2).
\end{align}

\begin{figure}[tbp!]
\centering
\includegraphics[width=8cm,clip]{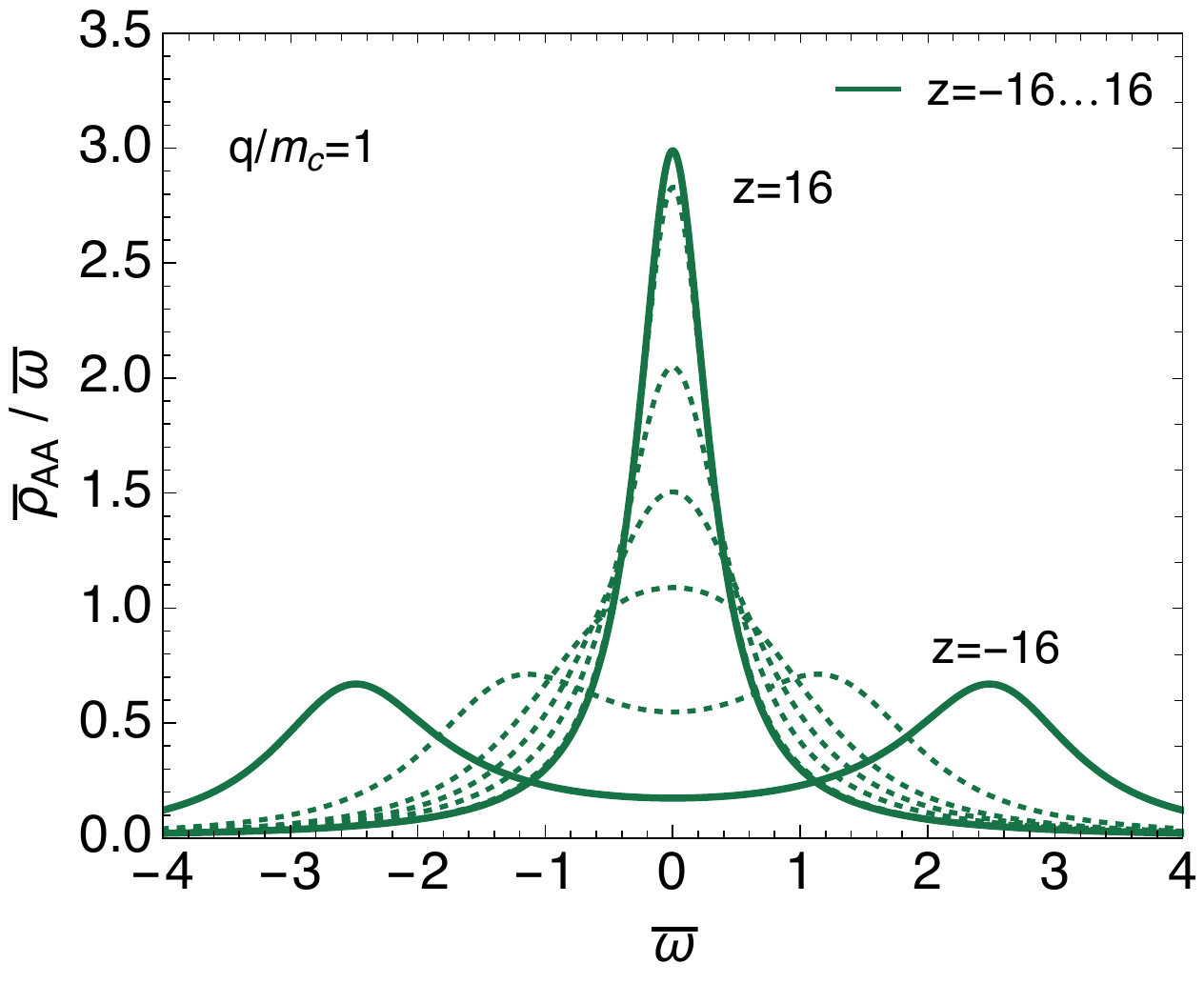}
\caption{The spectral density as a function of frequency at various values of $z$ above and below the phase transition. For $z\gg0$, $\rho_{AA}/\omega \propto Dk^2/(\omega^2 + (Dk^2)^2)$, which is the diffusive peak associated with quark diffusion. For $z\ll 0 $,  $\rho_{AA}\propto {\Gamma_k}/{((-\omega+\omega_k)^2+(\Gamma_k/2)^2})$, there are two distinct poles, which can be interpreted as a pair of propagating pions. For the choice of units, see \cite{Grossi:2021gqi}.}
\label{fig:prop}
\end{figure}
We can see the behavior of the propagator arising from \eqref{lin-eom} in Fig.~\ref{fig:prop}, where we plot the spectral density of the axial charge density, which is related to the propagator via
\begin{align}
(\chi_0\omega_k)^{-2}G_{\rm sym}^{\varphi\varphi}=\frac{T}{\omega} \rho_{AA}.
\end{align}
The figure neatly summarizes the changing degrees of freedom through the chiral phase transition: from diffusive quarks for $z\gg0 $, i.e. $T>T_c$, to soft pions for $z\ll 0$.

\section{Modification to transport coefficients: focus on shear viscosity}\label{sec:mod}

In this section, we outline how to compute the correction to the shear viscosity due to the pion modes. We will compute this using the Kubo formula
\begin{align}
    2 T \eta    = \int d^4x 
   \llangle \frac{1}{2} \{T^{xy} (t, {\bm x}), T^{xy}   (0, { \bm  0})  \}
   \rrangle\, ,
\end{align}
where the relevant component of the shear tensor \eqref{emt} in the linearized regime is given by
\begin{align}
\Delta T^{xy}&=  \frac14 \tr(\partial^x\Sigma\partial^y\Sigma^\dagger+\partial^y\Sigma\partial^x\Sigma^\dagger)
= {\partial^x\delta\sigma\partial^y\delta\sigma}
+{\bar \sigma^2\partial^x\varphi_a\partial^y\varphi_a} \, .
\end{align}
As we see, there are two contributions, one arising from the condensate and the other from the linearized pions.
Thus, it is expedient to break up the computation into two pieces:
\begin{align*}
\Delta\eta
\equiv {I_{\sigma\sigma}^{xy}}+I_{\varphi\varphi}^{xy}
\end{align*}

\begin{figure}
\centering \includegraphics[width=4cm]{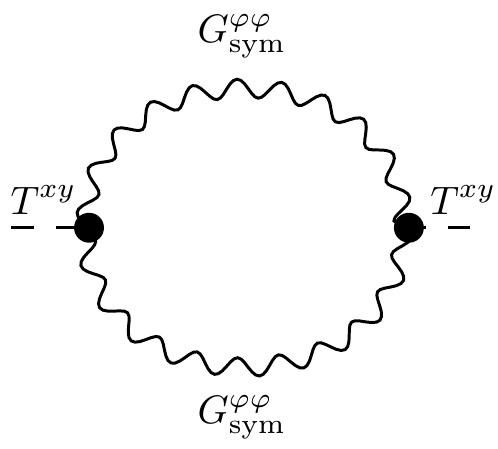}
\caption{A part of the diagram we need to compute for the shear viscosity.}
\label{fig:loop}
\end{figure}

\begin{figure}\label{fig:shear}
  \begin{center}
          {\includegraphics[width=0.7\textwidth]{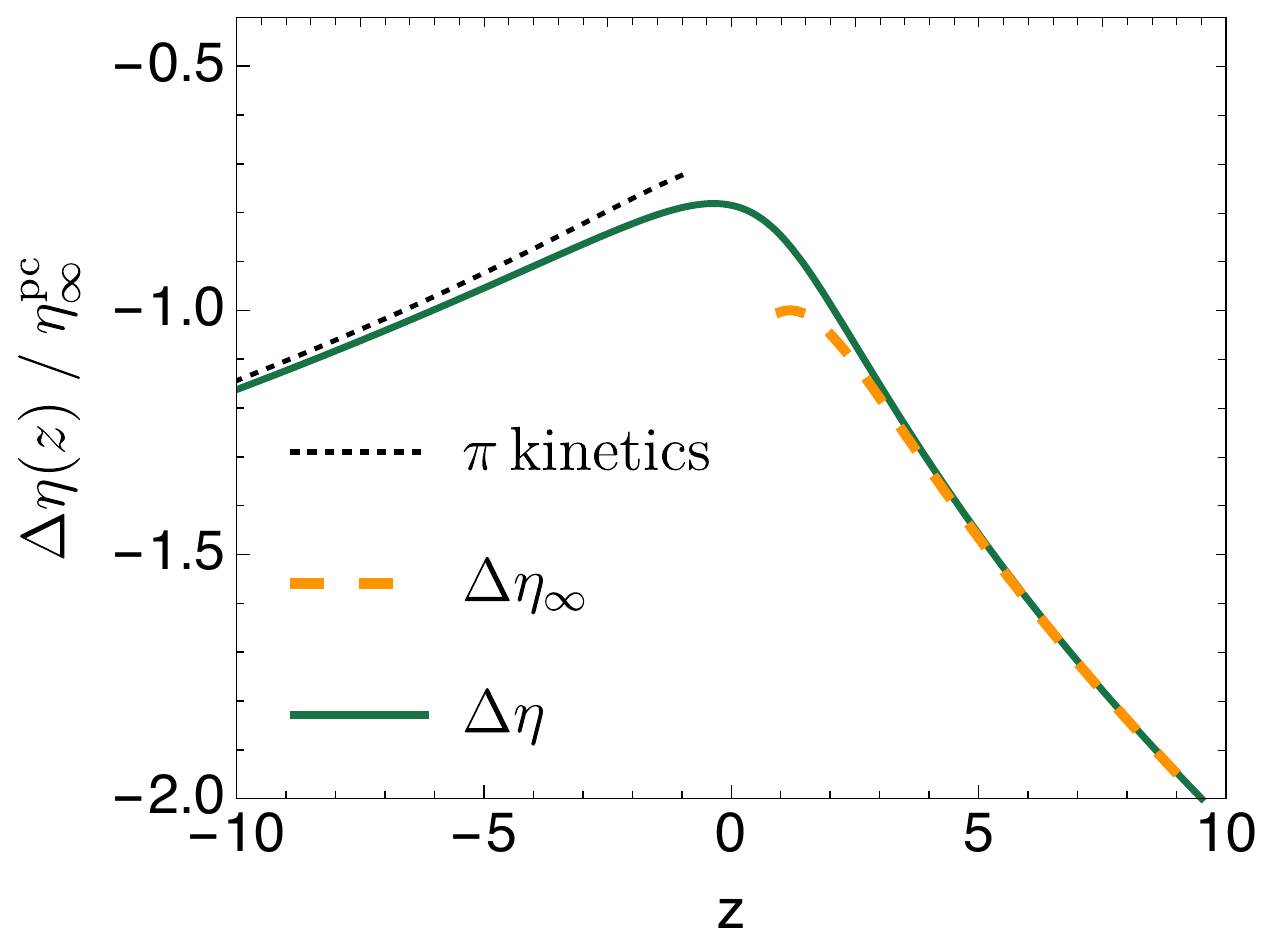}}

                        \caption{The correction to the shear viscosity, normalized by the value at the pseudocritical point at $z_{\rm pc}=1.19$, which we estimate to be $4\pi \eta^{\rm pc}_\infty/s=0.3$ \cite{Grossi:2021gqi}.}
    \end{center}
    \end{figure}

Computing the hydrodynamic loop dependent on the $\varphi$ propagator, $G^{\varphi\varphi}_{\rm sym},$ which is illustrated in Fig.~\ref{fig:loop}, we have
\begin{align}
{I_{\varphi\varphi}^{xy}}
=2 \int \frac{d^3  k}{(2\pi)^3}\frac{d  \omega}{(2\pi)} (k^x k^y G^{\varphi\varphi}_{sym})^2
\nonumber
 &={ \frac{2 T^2 }{30\pi^2 \Gamma}\Lambda}
  -{ \frac{ 2 T^2 m_{\sigma} }{32\pi \Gamma}}
  +\ldots
\end{align}
The ellipsis denotes subleading terms in the momentum cut-off, $\Lambda.$ Likewise, performing the integration dependent on the $\sigma$ propagator, we find
\begin{align}
I^{xy}_{\sigma\sigma}=\frac{2T^2}{30\pi^2\Gamma}\int_0^\Lambda \frac{k^6 dk}{(k^2+m_\sigma^2)^3}=\frac{T^2\Lambda}{15\pi^2\Gamma}-\frac{T^2m_\sigma}{16\pi\Gamma}+\ldots
\end{align}
Thus, we see that we have two pieces: a divergent term which can be absorbed into the definition of the hydrodynamic shear viscosity and a universal finite piece.
Thus, we are ready to state one of the main results of \cite{Grossi:2021gqi}: the shear viscosity gets a correction due to pions 
\begin{align}
\eta(z)=\eta_{\rm hydro}+{\Delta \eta}(z)\, .
\end{align}
The correction is plotted in Fig.~\ref{fig:shear}.  To understand the plot and the physics behind it, it is helpful to consider the two extreme limits. First, in the unbroken phase at large $z$, the propagator \eqref{propagator} simplifies dramatically and we have 
\begin{align}
\Delta \eta_\infty =-\frac{Tm_\sigma}{8\pi \Gamma},
\end{align}
which corresponds to the orange dotted line in the plot.

In the broken phase for $z\ll 0$, the relevant degrees of freedom are the soft pions. As such, one can compute the correction to the shear viscosity using another method, namely the Boltzmann equation in the relaxed time approximation for the pion distribution, $f_\pi$, in the effective fluid metric $G^\mn=-u^\mu u^\nu+v^2(x) \Delta^\mn$, where
$v^2$ is the  pion velocity, i.e. 
\begin{align}\frac{\partial \mathcal{H}}{\partial q_\mu } \frac{\partial f_\pi}{\partial x^\mu}-\frac{\partial \mathcal{H}}{\partial x^i }
 \frac{\partial f_\pi}{\partial q_i}=-\Gamma_q(\omega_q f_\pi -T),
 \end{align}
 where the effective Hamiltonian is given by $\mathcal{H}=\frac{1}{2}G^\mn q_\mu q_\nu+\frac{1}{2}v^2m^2$. 
The contributions of soft
pions to the transport coefficients of QCD in the broken phase were previously computed in \cite{Grossi:2020ezz}.

\section{Outlook to soft pion production}\label{sec:outlook}

As a look towards the future, in this section we will make an estimate for the soft pion enhancement near the critical point. As the system passes close to the chiral phase transition, the condensate and its phase are generated, resulting in the production of pions. These pions will have small momenta and could possibly imprint a signal of chiral dynamics in the detector.

We estimate the number of soft pions via the Bose-Einstein distribution $n=1/(e^{\omega(p)/T}-1)$. As such, we will need to provide the soft pion dispersion curve 
\begin{align}
\omega^2=v^2(p)p^2+m^2_{p}(p), 
\end{align}
which should interpolate between the two limiting dispersions: for soft pions with $q\ll \pi T$, 
\begin{align}
\omega_{q\ll \pi T}^2=v_0^2 p^2+m_{0p}^2,
\end{align}
while at larger momenta, we should recover the vacuum dispersion
\begin{align}
\omega_{\rm vac}^2= q^2+m_{\rm vac}^2.
\end{align}
The various parameters estimated at the pseudocritical point $z_{\rm pc}=1.19$ are as follows
\begin{align}\label{parameters}
v_0^2(z_{\rm pc})\simeq 0.25, \quad m_{0p}(z_{\rm pc})\simeq 0.1 \text{GeV}, \quad m_{\rm vac}^2=0.14 \text{GeV},
\end{align}
We model the interpolation via
\begin{align}
v^2(p)&=c^2(1-F(p/\Lambda)+v_0^2 F(p/\Lambda),\\
v^2(p)&=m_{\rm vac}^2(1-F(p/\Lambda)+m_{0p}^2 F(p/\Lambda),
\end{align}
where $F$ is any interpolating function, which is quadratic for small momenta and approaches zero when $p\sim \Lambda$. For our present purposes, we take
\begin{align}
F\left(\frac{p}{\Lambda}\right)=\frac{1}{1+\frac{1}{2}\left(\frac{p}{\Lambda}\right)^2+\left(\frac{p}{\Lambda}\right)^4}.
\end{align}

\begin{figure}[tbp!]
\centering
\begin{minipage}{\textwidth}
  
  \includegraphics[width=0.45\textwidth]{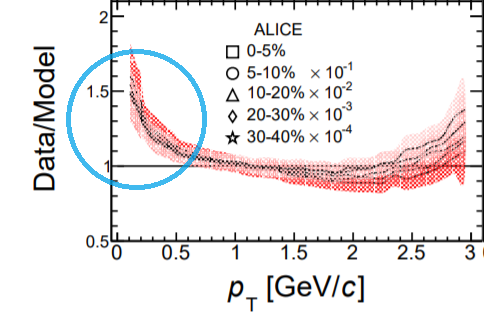}
    \includegraphics[width=0.45\textwidth]{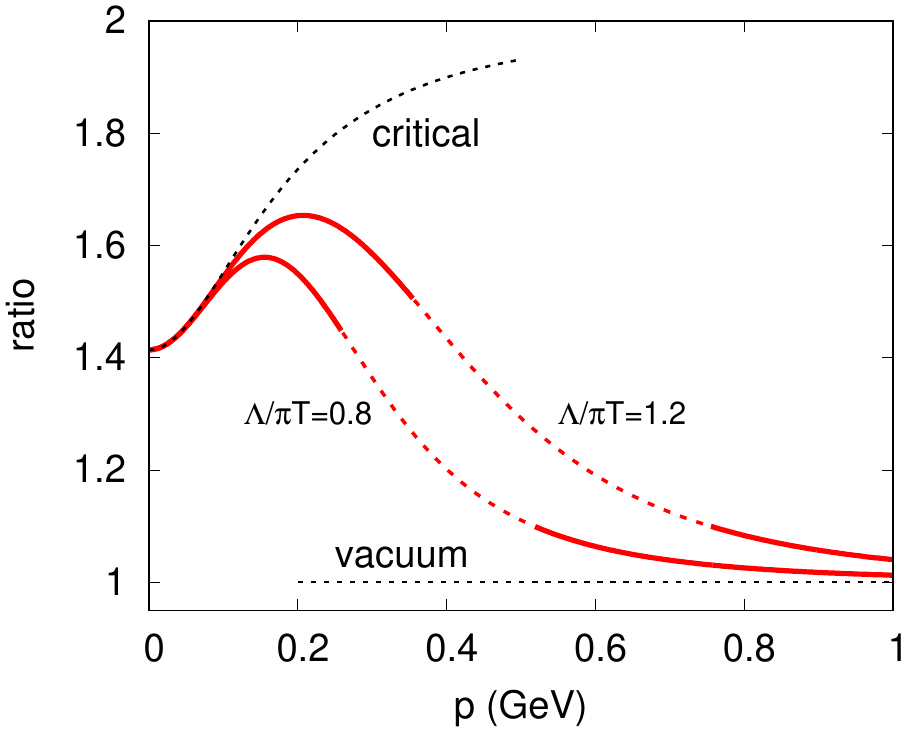}
  \caption{Left: comparison of experimental data to hydrodynamic model of pion yields from \cite{Devetak:2019lsk} with blue circle indicating the deviation for soft pions. Right: our estimate of soft pion yields using the model described in text.}
\label{fig:yields}
\end{minipage}%

\end{figure}

Now, we are ready to compute the ratio of the interpolation to the vacuum form, namely
\begin{align}
\text{ratio}=\frac{\frac{dN_{\rm crit}}{d^3p}}{\frac{dN_{\rm vac}}{d^3p}}=\frac{e^{\omega_{\rm vac}(p)/T}-1}{e^{\omega(p)/T}-1},
\end{align}
which we plot in Fig.~\ref{fig:yields} for a variety of cut-offs. 
The results are promising with the enhancement appearing in the same ballpark as the data. Of course, the model we outlined was rather crude and a more robust treatment is required, which would involve e.g. constraining dispersion via lattice, second order hydro, including resonance decays, etc. Moreover, since the parameters in \eqref{parameters} are Euclidean quantities, they can be measured more precisely on the lattice.
 In the future, there will be the upgrade to Inner Tracking System at ALICE \cite{alice}, which would allow experimentalists the chance to observe more low $p_T$ particles, especially soft pions.

%
%
%

%

\section*{Acknowledgments}
The author is supported by the Austrian Science Fund (FWF), project no. J4406.

\bibliography{vconf21-bib}

\begin{thebibliography}{13}

\bibitem{Jeon:2015dfa}
S.~Jeon, U.~Heinz, Int. J. Mod. Phys. E \textbf{24}, 1530010 (2015),
  \texttt{1503.03931}

\bibitem{HotQCD:2019xnw}
H.T. Ding et~al. (HotQCD), Phys. Rev. Lett. \textbf{123}, 062002 (2019),
  \texttt{1903.04801}

\bibitem{Kaczmarek:2020sif}
O.~Kaczmarek, F.~Karsch, A.~Lahiri, L.~Mazur, C.~Schmidt, \emph{{QCD phase
  transition in the chiral limit}} (2020), \texttt{2003.07920}

\bibitem{Grossi:2020ezz}
E.~Grossi, A.~Soloviev, D.~Teaney, F.~Yan, Phys. Rev. D \textbf{102}, 014042
  (2020), \texttt{2005.02885}

\bibitem{Grossi:2021gqi}
E.~Grossi, A.~Soloviev, D.~Teaney, F.~Yan, Phys. Rev. D \textbf{104}, 034025
  (2021), \texttt{2101.10847}

\bibitem{Rajagopal:1992qz}
K.~Rajagopal, F.~Wilczek, Nucl. Phys. B \textbf{399}, 395 (1993),
  \texttt{hep-ph/9210253}

\bibitem{Son:1999pa}
D.T. Son, Phys. Rev. Lett. \textbf{84}, 3771 (2000), \texttt{hep-ph/9912267}

\bibitem{Son:2002ci}
D.T. Son, M.A. Stephanov, Phys. Rev. D \textbf{66}, 076011 (2002),
  \texttt{hep-ph/0204226}

\bibitem{Florio:2021jlx}
A.~Florio, E.~Grossi, A.~Soloviev, D.~Teaney (2021), \texttt{2111.03640}

\bibitem{Engels:2009tv}
J.~Engels, O.~Vogt, Nucl. Phys. B \textbf{832}, 538 (2010), \texttt{0911.1939}

\bibitem{Engels:2011km}
J.~Engels, F.~Karsch, Phys. Rev. D \textbf{85}, 094506 (2012),
  \texttt{1105.0584}

\bibitem{alice}
\href{http://cds.cern.ch/record/2644611}{Expression of Interest for an ALICE ITS Upgrade in LS3}, (2018).
 
\bibitem{Devetak:2019lsk}
D.~Devetak, A.~Dubla, S.~Floerchinger, E.~Grossi, S.~Masciocchi,
  A.~Mazeliauskas, I.~Selyuzhenkov, JHEP \textbf{06}, 044 (2020),
  \texttt{1909.10485}

\end{thebibliography}
\bibliographystyle{woc}

\end{document}